\documentclass[sigconf]{acmart}
\hypersetup{
  colorlinks=true,
  linkcolor=blue
}

\usepackage{xcolor}
\usepackage{subfigure}

\definecolor{mycolor}{RGB}{127,185,87}

\usepackage{colortbl} % 提供 \columncolor 命令
\usepackage{xcolor}   % 提供颜色定义

\usepackage{multirow}
\usepackage[normalem]{ulem}
\useunder{\uline}{\ul}{}
\usepackage{enumitem}
\usepackage{epigraph} % 引言排版包
\AtBeginDocument{%
  }

\setcopyright{acmlicensed}
\copyrightyear{2026}
\acmYear{2026}
\acmDOI{XXXXXXX.XXXXXXX}
\acmConference[MM '26]{Proceedings of the 34th ACM International Conference on Multimedia}{November 10--14, 2026}{Rio de Janeiro, Brazil}
\acmSubmissionID{xxx}

\copyrightyear{2026}
\acmYear{2026}
\setcopyright{cc}
\setcctype{by-nc-nd}
\acmConference[MM '26]{Proceedings of the 34th ACM International Conference on Multimedia}{November 10--14, 2026}{Rio de Janeiro, Brazil}
\acmBooktitle{Proceedings of the 34th ACM International Conference on Multimedia (MM '26), November 10--14, 2026, Rio de Janeiro, Brazil}
\acmDOI{10.1145/3767308.3838604}
\acmISBN{979-8-4007-2213-4/2026/11}

\begin{document}

\title{MER 2026: From Discriminative Emotion Recognition \\to Generative Emotion Understanding}

\author{Zheng Lian}
\authornote{Corresponding author}
\affiliation{%
    \institution{Tongji University \& Institute of Automation, CAS}
    \city{Shanghai}
    \country{China}
}
% \email{lianzheng@tongji.edu.cn}
% 0000-0001-9477-0599

\author{Xiaojiang Peng}
\affiliation{%
    \institution{Shenzhen Technology University}
    \city{Shenzhen}
    \country{China}
}
% \email{pengxiaojiang@sztu.edu.cn}
% 0000-0002-5783-321X

\author{Kele Xu}
\affiliation{%
    \institution{National University of Defense Technology}
    \city{Changsha}
    \country{China}
}
% \email{kele.xu@ieee.org}
% 0000-0001-5997-5169

\author{Ziyu Jia}
\affiliation{%
    \institution{Institute of Automation, CAS}
    \city{Beijing}
    \country{China}
}
% \email{ziyu.jia@ia.ac.cn}
% 0000-0002-8523-1419

\author{Xinyi Che}
\affiliation{%
    \institution{Tongji University}
    \city{Shanghai}
    \country{China}
}
% \email{chexinyi@stu.scu.edu.cn}
% 0009-0008-7513-4501

\author{Kuofei Fang}
\affiliation{%
    \institution{Tongji University}
    \city{Shanghai}
    \country{China}
}
% \email{coffeyf@bupt.edu.cn}
% 0009-0007-7596-6140

\author{Zebang Cheng}
\affiliation{%
    \institution{Shenzhen University}
    \city{Shenzhen}
    \country{China}
}
% \email{zebang.cheng@gmail.com}
% 0009-0001-2854-7425

\author{Fei Ma}
\affiliation{%
    \institution{Guangdong Lab of Artificial Intelligence and Digital Economy (SZ)}
    \city{Shenzhen}
    \country{China}
}
% \email{mafei@gml.ac.cn}
% 0009-0002-5388-9125

\author{Laizhong Cui}
\affiliation{%
    \institution{Shenzhen University}
    \city{Shenzhen}
    \country{China}
}
% \email{cuilz@szu.edu.cn}
% 0000-0003-1991-290X

\author{Yazhou Zhang}
\affiliation{%
    \institution{Tianjin University}
    \city{Tianjin}
    \country{China}
}
% \email{yzhou_zhang@tju.edu.cn}
% 0000-0002-5699-0176

\author{Xin Liu}
\affiliation{%
    \institution{Shanghai Jiao Tong University}
    \city{Shanghai}
    \country{China}
}
% \email{linuxsino@gmail.com}
% 0000-0002-2242-6139

\author{Liang Yang}
\affiliation{%
    \institution{Dalian University of Technology}
    \city{Dalian}
    \country{China}
}
% \email{liang@dlut.edu.cn}
% 0000-0002-5557-7515

\author{Jia Li}
\affiliation{%
    \institution{Hefei University of Technology}
    \city{Hefei}
    \country{China}
}
% \email{jiali@hfut.edu.cn}
% 0000-0001-9446-249X

\author{Fan Zhang}
\affiliation{%
    \institution{The Chinese University of Hong Kong}
    \city{Hong Kong}
    \country{China}
}
% \email{fzhang25@cse.cuhk.edu.hk}
% 0000-0001-5250-7258

\author{Liumeng Xue}
\affiliation{%
    \institution{Nanjing University}
    \city{Nanjing}
    \country{China}
}
% \email{lmxue@nju.edu.cn}
% 0000-0003-2815-8494

\author{Erik Cambria}
\affiliation{%
    \institution{Nanyang Technological University}
    \country{Singapore}
}
% \email{cambria@sentic.net}
% 0000-0002-3030-1280

\author{Guoying Zhao}
\affiliation{%
    \institution{University of Oulu}
    \city{Oulu}
    \country{Finland}
}
% \email{guoying.zhao@oulu.fi}
% 0000-0003-3694-206X

\author{Björn W. Schuller}
\affiliation{%
    \institution{Technical University of Munich}
    \city{Munich}
    \country{Germany}
}
% \email{schuller@ieee.org}
% 0000-0002-6478-8699

\author{Jianhua Tao}
\authornotemark[1]
\affiliation{%
    \institution{Tsinghua University}
    \city{Beijing}
    \country{China}
}
% \email{jhtao@tsinghua.edu.cn}
% 0000-0002-0477-587X

\renewcommand{\shortauthors}{Zheng Lian et al.}
%% No italics, no superscripts, not anonymous
%% Use footnote or author note to identify equal contribution, shared contribution, and/or contact author info

\begin{abstract}

MER2026 marks the fourth edition of the MER series of challenges. The MER series provides valuable data resources to the research community and offers tasks centered on recent research trends, establishing itself as one of the largest challenges in the field. Throughout its history, the focus of MER has shifted from \emph{discriminative emotion recognition} to \emph{generative emotion understanding}. Specifically, MER2023 concentrated on \emph{discriminative emotion recognition}, restricting the emotion recognition scope to fixed basic labels. In MER2024 and MER2025, we transitioned to \emph{generative emotion understanding} and introduced two new tasks: fine-grained emotion recognition and descriptive emotion analysis, aiming to leverage the extensive vocabulary and multimodal understanding capabilities of Multimodal Large Language Models (MLLMs) to facilitate fine-grained and explainable emotion recognition. Building on this trajectory, MER2026 continues to follow these research trends and contains four tracks: \textbf{MER-Cross} shifts the focus from individual to dyadic interaction scenarios; \textbf{MER-FG} centers on fine-grained emotion recognition; \textbf{MER-Prefer} aims to predict human preferences regarding different emotion descriptions; \textbf{MER-PS} focuses on emotion recognition based on physiological signals. More details regarding the dataset and baselines are available at \href{https://zeroqiaoba.github.io/MER-Challenge/}{\color{cyan} https://zeroqiaoba.github.io/MER-Challenge/}.

\end{abstract}


% \keywords{Do, Not, Us, This, Code, Put, the, Correct, Terms, for,
% Your, Paper}
% %% A "teaser" image appears between the author and affiliation
% %% information and the body of the document, and typically spans the


\maketitle

\section{Introduction}
With the recent advancements in embodied AI, enabling robots to better understand human emotions and enhance their emotional intelligence has emerged as a significant research focus~\cite{arbib2004emotions, duan2022survey}. This development can facilitate their seamless integration into human society. Emotion is an internal human state expressed through various modalities, including audio, video, text, and physiological signals~\cite{plutchik1980general, picard2000affective}. This has spurred the growth of Multimodal Emotion Recognition (MER), which seeks to integrate diverse cues to achieve a more comprehensive understanding of human emotions~\cite{lian2026merbench}.

Given the significance of this task, we launched the MER series of challenges in 2023. MER2023@ACM Multimedia~\cite{lian2023mer} was our first organized challenge, aiming to advance the recognition of basic emotions through semi-supervised and multi-label learning techniques. MER2024@IJCAI~\cite{lian2024mer} shifted the focus from basic emotions to fine-grained emotions and introduced a new track dedicated to \emph{open-vocabulary emotions}~\cite{lian2025ov}. Building on this progression, MER2025@ACM Multimedia~\cite{lian2025mer} added another track, \emph{descriptive emotions}, which moves beyond simple emotion label prediction by emphasizing evidence-based emotion understanding~\cite{lian2023explainable,senticnet}. Reflecting on this developmental trajectory, MER challenges have transitioned from \emph{discriminative emotion recognition} to \emph{generative emotion understanding}: the former emphasizes basic emotion labels, whereas the latter prioritizes fine-grained, descriptive emotion representations. MER2026 represents the fourth installment of the MER challenge series. Aligned with current research trends, this year’s iteration introduces four tracks:

\textbf{Track 1. Interlocutor Emotion (MER-Cross).} This track inaugurates a shift from individual scenarios to dyadic interactions. Unlike prior works that focused on isolated speakers, MER-Cross targets the interlocutor's emotions. By integrating this focus with traditional tasks, we can capture the emotional states of both speakers in dyadic interactions. More details are illustrated in Figure~\ref{fig:cross}.

\textbf{Track 2. Fine-grained Emotion (MER-FG).} This track was initially proposed at MER2024, and this year marks our third time organizing it. Human emotions are relatively complex, extending far beyond basic labels. In this track, participants can predict any number of emotion labels across diverse categories, expanding the recognition scope from basic emotions to more nuanced ones. This task helps in better understanding the nuanced emotional states.

\textbf{Track 3. Emotion Preference (MER-Prefer).} This is a newly introduced track designed to predict human preferences toward different emotion descriptions. The concept of emotion preference was first introduced in EmoPrefer \cite{lian2026emoprefer}, playing a crucial role in training models capable of understanding human emotions.

\textbf{Track 4. Physiological Signal-Based Emotion (MER-PS).} Beyond human behavioral signals, emotions can also be recognized via physiological signals. In this track, we utilize synchronized EEG–fNIRS signals to predict dynamic emotions.

% Reflecting on the history of the MER challenges, participation has steadily increased from MER2023 to MER2025. To date, the MER dataset has been downloaded over 20K times, establishing it as one of the largest emotion recognition challenges in the research community. This year, we aim to further expand the influence and attract more participants to join the challenge.

\section{MER-Cross: Interlocutor Emotion}
\label{sec:mer_cross}

\begin{figure}[t]
	\centering
	\includegraphics[width=\linewidth]{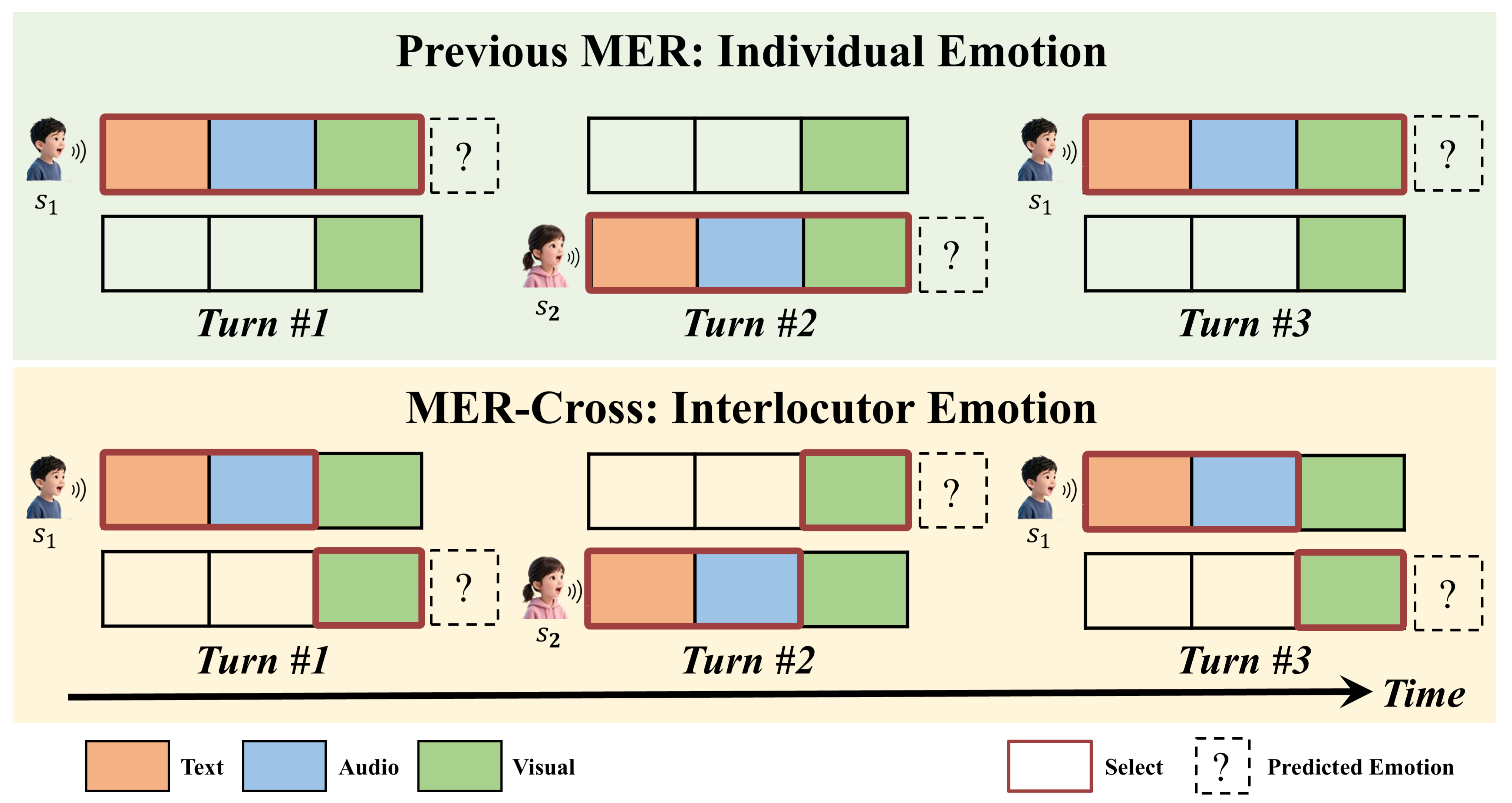}
	\caption{\textbf{MER-Cross}. Unlike previous tasks that focus on isolated speakers, we aim to predict the emotions of interlocutors. Thus, we can capture the emotions of both speakers.}
	\label{fig:cross}
\end{figure}

MER-Cross is a newly introduced track that shifts the focus from individual scenarios to dyadic interactions. As shown in Figure~\ref{fig:cross}, in such dyadic scenarios, each speaker talks in different turns. Specifically, consider two characters denoted as $s_1$ and $s_2$. In the first turn, $s_1$ is speaking, and multimodal clues are available for $s_1$, while $s_2$ is listening and only visual clues are accessible. Previous MER tasks focused primarily on individual emotions (i.e., the emotions of $s_1$). In contrast, MER-Cross shifts the focus to the interlocutor's emotions (i.e., the emotions of $s_2$). Therefore, by integrating traditional tasks with MER-Cross, we can capture the emotional states of both characters in dynamic interaction scenarios.

\subsection{Dataset}
\label{sec:mer_cross_dataset}
\paragraph{Sample Selection.} Taking turn \#1 in Figure~\ref{fig:cross} as an example, we require samples containing $s_1$'s audio and text information alongside $s_2$'s visual information. To this end, we employ TalkNet \cite{tao2021someone} for automatic active speaker detection and select samples with unmatched audio and video streams. However, manual inspection reveals that these automatic detections contain some errors. Meanwhile, there are some cases where a character seems not to speak, but the voice represents their inner thoughts, rather than coming from another speaker. Therefore, we implement an additional manual checking process to filter out these improper samples.

\paragraph{Annotation.} In our annotation process, we first conduct a preliminary assessment in which three annotators (undergraduate students from our institution) label 12 samples; annotators exhibiting significant discrepancies with the majority are filtered out. Then, we hire the filtered annotators to label the characters' emotional states and use the majority vote as the final label. Finally, we obtain 574 samples, with dataset statistics provided in Table~\ref{tab:statistic_cross}. In this track, we provide 9,395 samples with individual emotions for training, encouraging participants to investigate whether leveraging individual emotion data can enhance interlocutor emotion recognition.

\begin{table}[h]
	\centering
	\caption{Dataset statistics for MER-Cross. In MER-Cross, we provide 9,395 samples with individual emotions for training and 574 samples with interlocutor emotions for testing. %Participants can investigate whether leveraging individual emotion data can enhance interlocutor emotion recognition.
    }
	\label{tab:statistic_cross}
    \scalebox{0.9}{
	\begin{tabular}{l|cc|cc}
		\toprule
		   & \multicolumn{2}{c|}{\textbf{Train\&Val}} & \multicolumn{2}{c}{\textbf{Test}} \\
           & Type & \#Samples   & Type & \#Samples \\
          \midrule 
          MER-SEMI (2023) \cite{lian2023mer} & Individual  & 3,373 & Individual  & 834   \\ 
          MER-SEMI (2024) \cite{lian2024mer} & Individual  & 5,030 & Individual  & 1,169 \\ 
          MER-SEMI (2025) \cite{lian2025mer} & Individual  & 7,369 & Individual  & 2,026 \\ 
          MER-Cross (2026)                   & Individual  & 9,395 & Iterlocutor & 574   \\
		\bottomrule
	\end{tabular}
    }
\end{table}

\begin{table}[t]
	\centering
	\renewcommand\tabcolsep{2.4pt}
	\caption{Baseline results (\%) for MER-Cross. We report both unimodal and multimodal results. For the Train\&Val set, we present five-fold cross-validation results. We then select the best-performing model based on the Train\&Val set and report its performance on the test set. %The grey column indicates the evaluation metric used for final ranking.
    }
	\label{tab:baseline_cross}
    \scalebox{0.9}{
	\begin{tabular}{l|cc|>{\columncolor[gray]{0.9}}cc}
		\toprule
		\multirow{2}{*}{\textbf{Features}} & \multicolumn{2}{c|}{\textbf{Train$\&$Val}} & \multicolumn{2}{c}{\textbf{Test}} \\
		& WAF $(\uparrow)$ & ACC $(\uparrow)$ & WAF $(\uparrow)$ & ACC $(\uparrow)$ \\
		
		\midrule
		\multicolumn{5}{c}{\emph{Acoustic Modality}} \\
		\midrule
        
WavLM-base        & 58.44$\pm$0.09 & 58.83$\pm$0.10 & 30.14$\pm$0.50 & 32.70$\pm$0.34 \\
wav2vec 2.0-base  & 67.38$\pm$0.27 & 67.52$\pm$0.28 & 31.84$\pm$0.28 & 33.40$\pm$0.23 \\
wav2vec 2.0-large & 67.64$\pm$0.42 & 67.77$\pm$0.47 & 32.63$\pm$0.24 & 34.25$\pm$0.21 \\
HUBERT-large      & 76.51$\pm$0.15 & 76.60$\pm$0.15 & 35.25$\pm$0.16 & 36.60$\pm$0.15 \\
HUBERT-base       & 72.78$\pm$0.08 & 72.85$\pm$0.11 & 35.74$\pm$0.41 & 37.26$\pm$0.32 \\

		\midrule
		\multicolumn{5}{c}{\emph{Lexical Modality}} \\
		\midrule

RoBERTa-base  & 53.63$\pm$0.16 & 53.82$\pm$0.16 & 31.93$\pm$0.40 & 32.23$\pm$0.34 \\
RoBERTa-large & 54.61$\pm$0.22 & 54.73$\pm$0.22 & 32.19$\pm$0.51 & 32.70$\pm$0.33 \\
MacBERT-base  & 53.42$\pm$0.13 & 53.58$\pm$0.14 & 32.24$\pm$0.36 & 32.27$\pm$0.20 \\
MacBERT-large & 54.14$\pm$0.12 & 54.20$\pm$0.09 & 33.68$\pm$0.25 & 33.95$\pm$0.39 \\

		\midrule
		\multicolumn{5}{c}{\emph{Visual Modality}} \\
		\midrule
        
VideoMAE-large & 56.66$\pm$0.21 & 57.34$\pm$0.14 & 41.08$\pm$0.71 & 45.89$\pm$0.35 \\
ResNet-FER2013 & 56.57$\pm$0.17 & 57.88$\pm$0.20 & 45.86$\pm$0.54 & 48.03$\pm$0.59 \\
EmoNet         & 52.17$\pm$0.34 & 53.58$\pm$0.30 & 47.03$\pm$0.68 & 49.95$\pm$0.56 \\
SENet-FER2013  & 57.01$\pm$0.16 & 58.23$\pm$0.23 & 51.55$\pm$0.41 & 51.92$\pm$0.58 \\
CLIP-base      & 60.86$\pm$0.25 & 61.83$\pm$0.29 & 52.43$\pm$0.41 & 55.16$\pm$0.28 \\
CLIP-large     & 65.28$\pm$0.13 & 65.84$\pm$0.14 & 58.88$\pm$0.43 & 60.61$\pm$0.44 \\

        \midrule
		\multicolumn{5}{c}{\emph{Acoustic + Lexical + Visual}} \\
		\midrule
        
Top1 & 80.91$\pm$0.10 & 80.98$\pm$0.10 & 57.44$\pm$0.25 & 57.94$\pm$0.33 \\
Top2 & 82.01$\pm$0.12 & 82.07$\pm$0.13 & 55.24$\pm$0.32 & 55.87$\pm$0.51 \\

		\bottomrule

	\end{tabular}
    }
\end{table}

\subsection{Evaluation Metric}
This track focuses on basic emotion recognition, where predictions are limited to six predefined categories: \emph{neutral}, \emph{anger}, \emph{happiness}, \emph{sadness}, \emph{worry}, and \emph{surprise}. Following previous work~\cite{lian2025mer,lian2026merbench}, we adopt the weighted F1-score (WAF) as the primary evaluation metric. Additionally, we report accuracy (ACC) as a secondary metric.

\subsection{Baselines}
Table~\ref{tab:baseline_cross} presents both unimodal and multimodal results. For the multimodal experiments, we selected the best-performing unimodal features for fusion. Specifically, ``Top-1'' denotes the selection of the best unimodal feature per modality (i.e., \emph{HUBERT-base, MacBERT-large, and CLIP-large}); ``Top-2'' indicates the use of the top two features per modality (i.e., \emph{HUBERT-large, HUBERT-base, MacBERT-base, MacBERT-large, CLIP-base, and CLIP-large}). To mitigate the impact of randomness, we ran each experiment multiple times and reported the average scores accompanied by standard deviations. Experimental results reveal a considerable performance gap between individual and interlocutor emotions. For instance, while acoustic features achieve the best performance for individual emotions, they perform poorly for interlocutor emotions. Furthermore, although multimodal fusion yields noticeable performance gains for individual emotions, it does not improve performance on interlocutor emotions. Therefore, participants are encouraged to develop novel strategies to bridge this gap.

\section{MER-FG: Fine-grained Emotion}
\label{sec:mer_fg}
\begin{figure}[t]
	\centering
	\includegraphics[width=\linewidth]{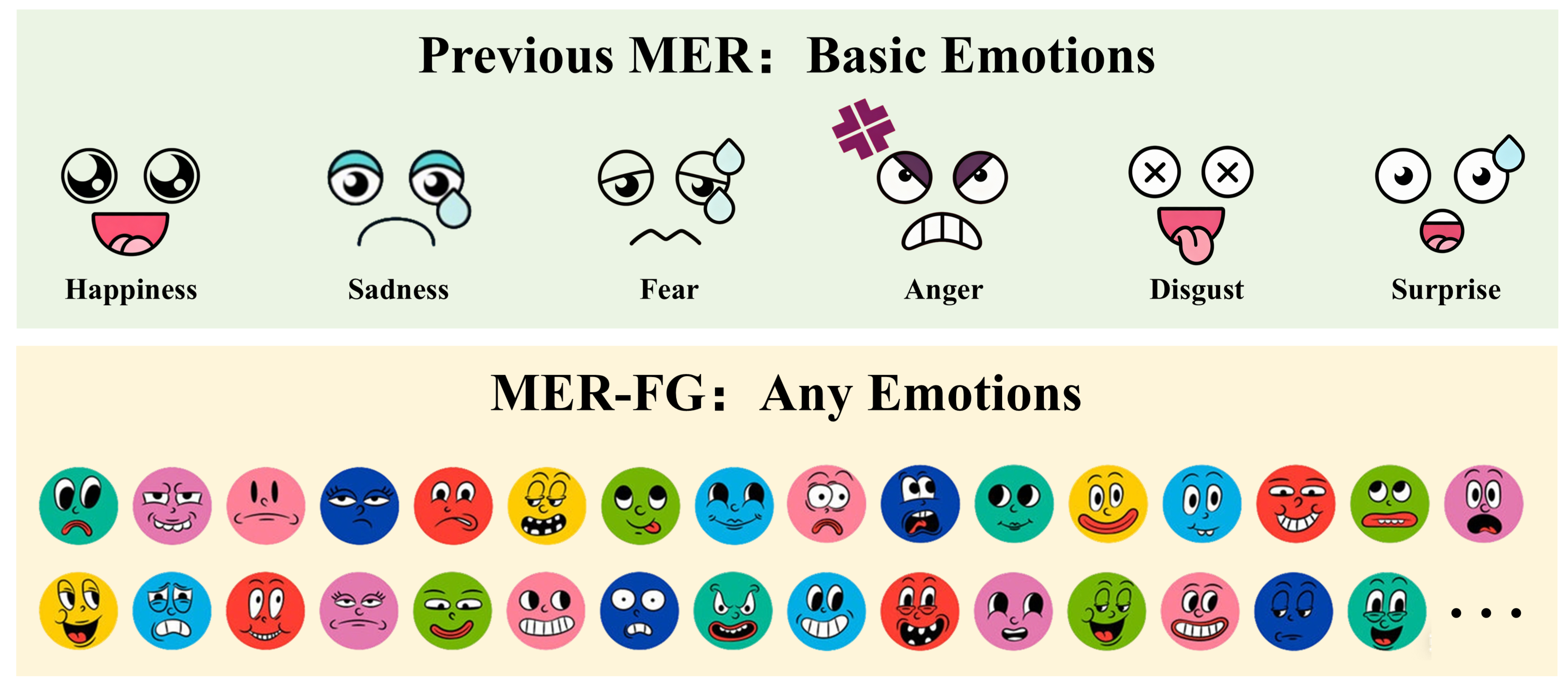}
	\caption{\textbf{MER-FG}. Unlike previous MER tasks that focus on basic emotions, MER-FG extends the recognition scope to encompass any emotion category.}
	\label{fig:fg}
\end{figure}

MER-FG was first introduced at MER2024~\cite{lian2024mer}, and this year marks its third edition. The primary motivation behind MER-FG is that human emotion encompasses a vast array of emotion words, extending far beyond basic emotions~\cite{ekman1992there}. Confining such rich emotional expression within a fixed label set can result in inaccurate emotion representation. Therefore, MER-FG aims to go beyond basic emotions and enable fine-grained emotion recognition~\cite{lian2025ov}. Figure~\ref{fig:fg} illustrates the differences between MER-FG and previous tasks.

\subsection{Dataset}

\paragraph{Challenge Tendency.} Table~\ref{tab:statistic_fg} summarizes the dataset statistics for MER-FG across previous challenges and the current edition. In MER 2024, we utilized OV-MERD for training and provided an additional 200 manually labeled samples for testing. In MER 2025, we introduced the large-scale, automatically labeled MER-Caption+ dataset for training and expanded the test set to 1,200 samples. For this year's challenge, we merged OV-MERD with the MER2025-FG test set to form the new Human-OV dataset, while also providing an additional 1,000 samples for testing.

\paragraph{Annotation.} We follow the annotation pipeline in previous work \cite{lian2025mer}. Specifically, we utilize the submission results from MER2025-FG, where participating teams were required to predict open-vocabulary labels for 20,000 samples. We merge the predictions from the top-performing teams to construct candidate label sets for each sample. Subsequently, we randomly select 1,000 samples for a two-round manual validation process. In the first round, we retain all labels selected by at least one annotator to ensure completeness; in the second round, we preserve only the majority-voted labels to ensure accuracy. For this purpose, we recruit five annotators (master’s students from our Dataset Chairs’ Group) and randomly assign each sample to three annotators. Finally, we obtain an additional 1,000 manually labeled samples for testing.

\begin{table}[t]
	\centering
	\caption{Dataset statistics for MER-FG. This year, we provide two training datasets: Human-OV with manual labels and MER-Caption+ with automatic labels. Additionally, we provide 1,000 manually labeled samples for testing.}
	\label{tab:statistic_fg}
    \scalebox{0.9}{
	\begin{tabular}{l|cc|c}
		\toprule
        & \multicolumn{2}{c|}{\textbf{Train\&Val}} & \textbf{Test} \\
        & Dataset & \#Samples & \\
        \midrule 
        MER2024-FG~\cite{lian2024mer} & OV-MERD \cite{lian2023mer} & 332 & 200   \\
        \midrule 
        \multirow{2}{*}{MER2025-FG~\cite{lian2025mer}} & OV-MERD \cite{lian2023mer} & 332 & \multirow{2}{*}{1,200} \\
                                      & MER-Caption+ \cite{lian2025affectgpt} & 31,327 & \\
        \midrule 
        \multirow{2}{*}{MER2026-FG}   & Human-OV & 1,532 & \multirow{2}{*}{1,000} \\
                                      & MER-Caption+ \cite{lian2025affectgpt} & 31,327 & \\
		\bottomrule
	\end{tabular}
    }
\end{table}

\subsection{Evaluation Metric}

This relaxed label space imposes new requirements on evaluation metrics. In this track, we utilize emotion wheel-based (EW-based) metrics for the final ranking, following official guidelines \cite{lian2025affectgpt,lian2026affectgptr1}.

\textbf{Grouping.} Since we do not restrict the label space, the model may generate synonyms. Following prior works \cite{lian2025affectgpt}, we use grouping information from emotion wheels to mitigate the effects of synonymy. Specifically, we define a three-level grouping strategy:

{(L1)} We normalize different forms of emotion words to their base form. This function is denoted as $A(\cdot)$.

{(L2)} We map synonyms to a unified label. For example, we map \emph{happy} and \emph{joyful} to \emph{happy}. We denote this function as $B(\cdot)$.

{(L3)} In emotion wheels, basic emotions reside in the innermost layer, while more nuanced labels are arranged in the outer layers. Following prior works \cite{lian2025affectgpt}, we employ five emotion wheels $\{w_i\}_{i=1}^5$. For each wheel $w_i$, we define a mapping function $C_{i}(\cdot)$ that transforms labels from outer layers to their corresponding innermost layer. These three-level grouping functions can be summarized as:
\begin{equation}
\mathcal{F}_i(\cdot) = C_i{\left(B\left(A\left(\cdot\right)\right)\right)}, i \in [1, 5].
\end{equation}

\textbf{Metric.} Let the dataset consist of $N$ samples. For sample $x_j$, the true labels are denoted as $\mathbf{Y}_j$ and the predicted labels are denoted as $\mathbf{\hat{Y}}_j$. We use the average F-scores across five wheels for ranking:
\begin{equation}
\mbox{Precision}_{\mbox{s}}^i  = \frac{1}{N}\sum_{j=1}^{N}\frac{\left|\mathcal{F}_i( \mathbf{Y}_j ) \cap \mathcal{F}_i(\mathbf{\hat{Y}}_j)\right|}{\left|\mathcal{F}_i(\mathbf{\hat{Y}}_j)\right|},
\end{equation}
\begin{equation}
\mbox{Recall}_{\mbox{s}}^i = \frac{1}{N}\sum_{j=1}^{N}\frac{\left|\mathcal{F}_i( \mathbf{Y}_j ) \cap \mathcal{F}_i(\mathbf{\hat{Y}}_j)\right|}{\left|\mathcal{F}_i(\mathbf{{Y}}_j)\right|},
\end{equation}
\begin{equation}
\mbox{F}_{\mbox{s}}^i  = 2\times\frac{\mbox{Precision}_{\mbox{s}}^i \times\mbox{Recall}_{\mbox{s}}^i }{\mbox{Precision}_{\mbox{s}}^i +\mbox{Recall}_{\mbox{s}}^i },
\end{equation}
\begin{equation}
\mbox{Metric} = \text{Avg}[\mbox{F}_{\mbox{s}}^{i} ], i \in [1, 5].
\end{equation}

\subsection{Baselines}
Table~\ref{tab:baseline_fg} presents the performance of both zero-shot baselines and post-training models. For the zero-shot setting, no additional post-training is conducted; for the post-training setting, we adopt AffectGPT as the base model and perform further training. We observe that post-training models achieve noticeable performance gains over zero-shot baselines, primarily because they are specifically optimized for open-vocabulary emotion recognition. Furthermore, among the post-training models, there are no considerable differences between those trained on Human-OV versus MER-Caption+. This suggests a trade-off between data quality and quantity: Human-OV contains fewer samples but offers higher-quality labels than MER-Caption+. Consequently, increasing data quantity or improving data quality can both lead to better performance.

\begin{table}[t]
	\centering
	\caption{Baseline results (\%) for MER-FG. We report the performance of both zero-shot baselines and post-training models. For the zero-shot setting, no additional post-training is conducted; for the post-training setting, we adopt AffectGPT as the backbone and perform additional training.}
	\label{tab:baseline_fg}
    \scalebox{0.9}{
	\begin{tabular}{l|c>{\columncolor[gray]{0.9}}c}
		\toprule
		  \textbf{Model} & \textbf{\textbf{Post-Training Data}} & \textbf{Test Performance} \\

\midrule
\multicolumn{3}{c}{\emph{Zero-shot Baselines}} \\
\midrule

Video-LLaVA   & --  & 30.83 \\
Qwen-Audio    & --  & 34.90 \\
Video-ChatGPT & --  & 36.51 \\
Chat-UniVi    & --  & 43.03 \\
LLaMA-VID     & --  & 44.34 \\
SALMONN       & --  & 47.38 \\

\midrule
\multicolumn{3}{c}{\emph{Post-training Models}} \\
\midrule
        
AffectGPT & Human-OV     & 59.54 \\
AffectGPT & MER-Caption+ & 60.27 \\

		\bottomrule

	\end{tabular}
    }
\end{table}

\section{MER-Prefer: Emotion Preference}
\label{sec:mer_prefer}
\begin{figure}[t]
	\centering
	\includegraphics[width=\linewidth]{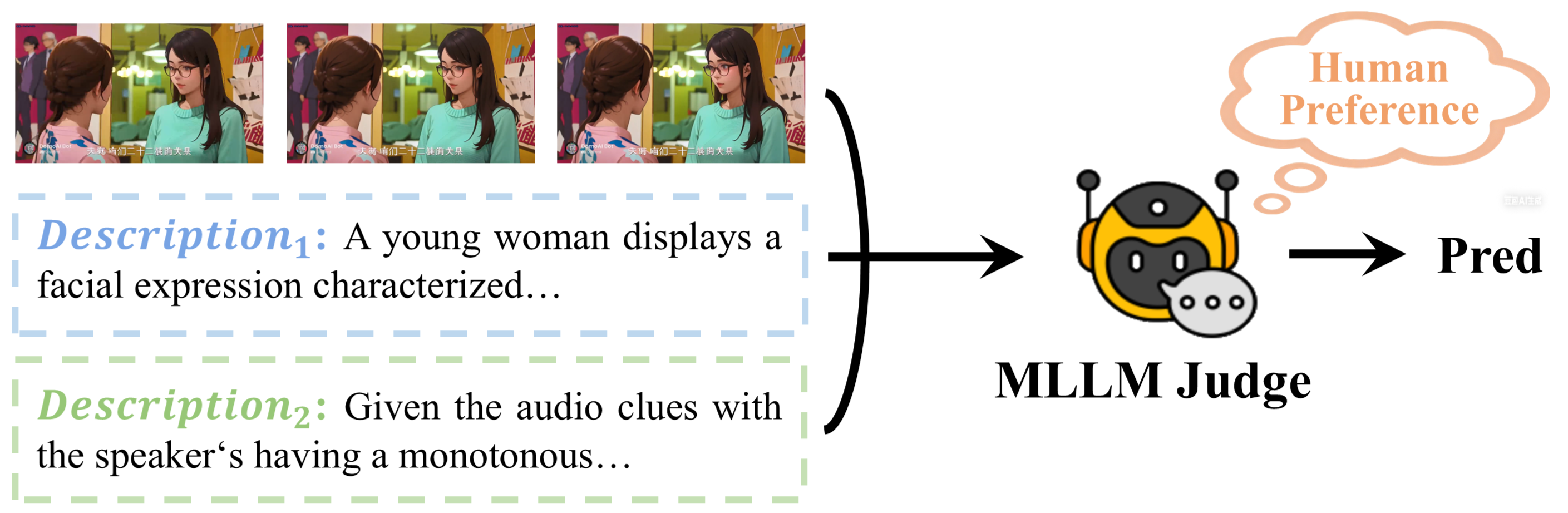}
	\caption{\textbf{MER-Prefer}. Given a video, the model needs to determine which emotion description is preferred by humans.}
	\label{fig:prefer}
\end{figure}

MER-Prefer is a newly introduced track. The concept of emotion preference was first proposed in EmoPrefer~\cite{lian2026emoprefer}, which predicts, for a given video, which of two emotion descriptions is preferred by human annotators. This task plays a crucial role in training reward models capable of understanding human emotions. Figure~\ref{fig:prefer} illustrates the input and output for this track.

\subsection{Dataset}
We provide two preference datasets, EmoPrefer-Data \cite{lian2026emoprefer} and EmoPrefer-Data-V2 \cite{lian2026emoprefer}, and summarize their statistics in Table~\ref{tab:statistic_prefer}. The primary distinction is that the former utilizes majority-voted preferences, while the latter relies on single-annotator preferences. For the test set, we collected additional samples. Specifically, we hired five annotators who had passed a preliminary qualification test to ensure annotation quality (aligned with Section~\ref{sec:mer_cross_dataset}). We then collected 1,000 samples, denoted as $\{(x_n, d_1^n, d_2^n)\}_{i=1}^N$, where $x_n$ is a video and $d_1^n$ and $d_2^n$ are the corresponding descriptions. For each sample, three annotators were randomly selected to choose their preferred description. Annotations fell into one of three categories: (1) $d_1^n$ is preferred, (2) $d_2^n$ is preferred, or (3) \emph{tie}. Finally, only samples with consensus were retained, resulting in 380 samples. Since the \emph{tie} category is ambiguous \cite{lian2026emoprefer}, we removed the samples labeled as such, ultimately obtaining a test set of 379 samples.

\subsection{Evaluation Metric}
We adopt the official evaluation metric~\cite{lian2026emoprefer} for performance assessment. Specifically, we focus on two-class preference classification, where the task is to determine whether $d_1^n$ or $d_2^n$ is preferred. We employ the weighted F1-score (WAF) as the primary metric and accuracy (ACC) as the secondary metric. These metrics quantify the consistency between model predictions and human preferences.

\subsection{Baselines}
Table~\ref{tab:baseline_prefer} presents the performance of zero-shot baselines on MER-Prefer. We observe that the multimodal model, Qwen2.5-Omni, achieves the best performance. This suggests a strong correlation between human preferences and the multimodal cues inherent in videos. Consequently, integrating all modalities is essential for accurately capturing human preferences between different descriptions.

\begin{table}[t]
	\centering
	\caption{\textbf{Dataset statistics for MER-Prefer.} %We provide two training datasets: EmoPrefer-Data with majority-voted labels and EmoPrefer-Data-V2 with single-annotator labels. For testing, we provide 379 samples with majority-voted labels.
    }
	\label{tab:statistic_prefer}
    \scalebox{0.9}{
	\begin{tabular}{lcc|cc}
		\toprule
        \multicolumn{3}{c|}{\textbf{Train\&Val}} & \multicolumn{2}{c}{\textbf{Test}} \\
        Dataset & Label & \#Samples  & Label & \#Samples  \\
        \midrule 
        {EmoPrefer-Data}~\cite{lian2026emoprefer} & Major & 574 & \multirow{2}{*}{Major} & \multirow{2}{*}{379} \\
        {EmoPrefer-Data-V2}~\cite{lian2026emoprefer} & Single & 2,096 & & \\
		\bottomrule
	\end{tabular}
    }
\end{table}

\begin{table}[t]
	\centering
	\caption{Baseline results (\%) for MER-Prefer. We report the performance of zero-shot baselines. %The grey column indicates the evaluation metric used for final ranking.}
    }
	\label{tab:baseline_prefer}
    \scalebox{0.9}{
	\begin{tabular}{l|>{\columncolor[gray]{0.9}}cc}
		\toprule
		  \multirow{2}{*}{\textbf{Model}} & \multicolumn{2}{c}{\textbf{Test}} \\
                       & WAF $(\uparrow)$ & ACC $(\uparrow)$ \\

\midrule

Qwen2-Audio & 36.08 & 49.08 \\
Video-LLaVA  & 36.76 & 53.03 \\
LLaMA-VID  & 36.76 & 53.03 \\
LLaVA-Next-Video  & 41.31 & 55.15 \\
Qwen2.5-VL  & 76.77 & 77.84 \\
Qwen2.5-Omni  & 78.74 & 78.89 \\

		\bottomrule

	\end{tabular}
    }
\end{table}

\section{MER-PS: Physiological Signal-Based Emotion}
\label{sec:mer_ps}

MER-PS shifts multimodal emotion recognition from observable behavioral signals to physiological evidence, with a particular emphasis on brain activity. Compared with conventional behavioral cues, physiological signals provide a more direct window into the internal affective state and are less dependent on explicit facial or vocal expressions. To support this track, we introduce MER-PS, which focuses on predicting continuous affective dynamics from synchronized EEG and fNIRS signals. This track extends the scope of MER from external multimodal behaviors to internal physiological responses, and opens up new opportunities for studying emotion recognition under more implicit and fine-grained settings.

\subsection{Dataset}
\label{sec:mer_ps_dataset}

\begin{figure}[t]
	\centering
	\includegraphics[width=\linewidth]{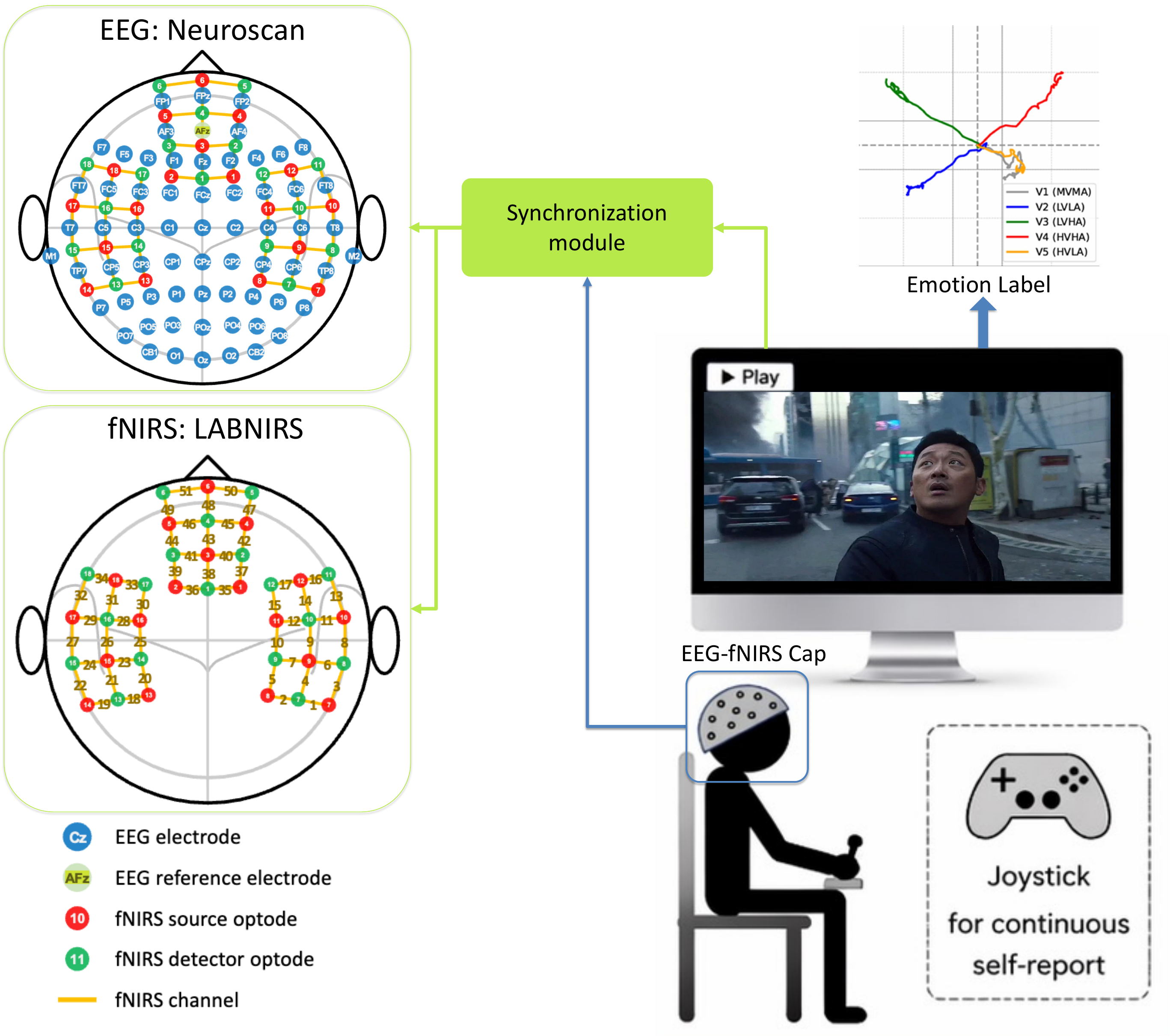}
	\caption{\textbf{MER-PS}. EEG and fNIRS signals are synchronously recorded during emotion elicitation, while participants continuously report their affective states via joystick-based valence--arousal annotation. %This track enables physiological signal-based continuous emotion prediction with temporally aligned multimodal recordings and dynamic labels.}
    }
	\label{fig:mer_ps_framework}
\end{figure}

\paragraph{Data Collection.}
The dataset used in this track is a synchronized EEG--fNIRS emotion dataset collected under realistic emotion-elicitation scenarios. Specifically, 30 participants watched 15 emotion-eliciting video clips, together with baseline sessions recorded before formal stimulation. During the experiment, EEG and fNIRS signals were synchronously acquired, enabling the joint observation of neuroelectrical activity and hemodynamic responses associated with human emotions. The overall recording and annotation framework of MER-PS is illustrated in Figure~\ref{fig:mer_ps_framework}. In addition to physiological recordings, the dataset also provides continuous affect annotations as well as post-trial self-assessment labels, forming a comprehensive benchmark for physiological signal-based emotion analysis.

\paragraph{Signal and Annotation.}
The MER-PS dataset contains 64-channel EEG signals recorded at 1000 Hz and 51-channel fNIRS signals recorded at 47.62 Hz, covering cortical regions that are closely related to affective processing. Specifically, the EEG signals were acquired using an ESI Neuroscan 64-channel wet-electrode system, with electrodes positioned according to the international 10--10 EEG system and the default reference electrode placed at AFz. The fNIRS signals were recorded using a Shimadzu LABNIRS system with three near-infrared wavelengths (780, 805, and 830 nm). In our recordings, the fNIRS optodes were arranged to cover three major cortical regions, i.e., the prefrontal lobes, the left temporal lobes, and the right temporal lobes. To support synchronized multimodal acquisition, an EEG--fNIRS joint cap and a signal synchronization module developed by FiStar were used to achieve joint acquisition and precise temporal alignment between the two modalities. A key characteristic of this dataset is that it provides real-time dynamic emotion annotations. During video viewing, each participant continuously reported their affective state through joystick-based valence and arousal trajectories sampled at 1 Hz, where each time step is associated with a two-dimensional affect label. In addition to these dynamic annotations, the dataset also includes post-trial self-assessments, which can serve as supplementary references for affective analysis. In this track, we focus on the continuous affect prediction setting and require participants to estimate the valence--arousal trajectory from synchronized physiological signals.

\subsection{Evaluation Metric}

We formulate emotion recognition as a regression problem over continuous affective dynamics. Given synchronized EEG-fNIRS signals, we predict the valence and arousal values at each time step:
\begin{equation}
\hat{\mathbf{y}}_{1:T}=f\left(\mathbf{x}^{\text{EEG}}_{1:L_e}, \mathbf{x}^{\text{fNIRS}}_{1:L_f}\right), \quad
\mathbf{y}_t=[v_t,a_t],
\end{equation}
where $\mathbf{x}^{\text{EEG}}$ and $\mathbf{x}^{\text{fNIRS}}$ denote the EEG and fNIRS time series, respectively, and $\mathbf{y}_t$ denotes the affect label at time step $t$. In the original annotation space, both valence and arousal are recorded at 1 Hz as integer values in $[1,255]$, with the neutral point centered at $(128,128)$. Moreover, the emotion-eliciting materials are designed to cover five targeted affective tendencies, namely high valence--high arousal, high valence--low arousal, low valence--high arousal, low valence--low arousal, and medium valence--medium arousal, thereby providing broad coverage in the valence--arousal space.

Following the official setting of this track, we normalize both valence and arousal labels into $[0,1]$ and evaluate the regression performance using the mean absolute error (MAE):
\begin{equation}
\text{MAE}=\frac{1}{N}\sum_{n=1}^{N}\left|y_n-\hat{y}_n\right|.
\end{equation}
To provide a unified ranking score, we use the average MAE over the two affect dimensions:
\begin{equation}
\text{Score}=\frac{\text{MAE}_{v}+\text{MAE}_{a}}{2},
\end{equation}
where $\text{MAE}_{v}$ and $\text{MAE}_{a}$ denote the mean absolute errors for valence and arousal. A lower score indicates better performance.

\subsection{Baselines}

Table~\ref{tab:baseline_ps} reports the baseline results for MER-PS. In this track, we adopt two baseline models, i.e., EEGNet~\cite{lawhern2018eegnet} and ASAC-Net~\cite{FANG2026104329}, and evaluate them under three input settings: EEG-only, fNIRS-only, and EEG+fNIRS. Specifically, EEGNet serves as a classical baseline for physiological signal modeling, while ASAC-Net represents a recent EEG--fNIRS-based method for emotion analysis. In this way, we can examine both the effect of model architecture and the contribution of multimodal physiological fusion.

Experimental results provide initial benchmarks for this track. Since studies on continuous affect prediction from synchronized EEG and fNIRS signals are still limited, especially for effective bimodal collaboration, participants are encouraged to explore novel strategies for EEG--fNIRS coordination, such as temporal alignment, cross-modal interaction, correlation modeling, and fusion learning, in order to achieve better performance in MER-PS.

\begin{table}[t]
	\centering
	\renewcommand\tabcolsep{4.5pt}
	\caption{Baseline results for MER-PS. %We report the regression performance for valence and arousal prediction, together with the final ranking score. The grey column indicates the evaluation metric used for final ranking.}
    }
	\label{tab:baseline_ps}
    \scalebox{0.95}{
	\begin{tabular}{l|l|cc|>{\columncolor[gray]{0.9}}c}
		\toprule
		\multirow{2}{*}{\textbf{Architecture}} & \multirow{2}{*}{\textbf{Features}} & \multicolumn{2}{c|}{\textbf{Test}} & \textbf{Test} \\
		&  & MAE$_v$ $(\downarrow)$ & MAE$_a$ $(\downarrow)$ & Score $(\downarrow)$ \\
		\midrule
		\multirow{3}{*}{EEGNet~\cite{lawhern2018eegnet}}
		& EEG-only      & 0.2682 & 0.2319 & 0.2501 \\
		& fNIRS-only    & 0.2567 & 0.2245 & 0.2406 \\
		& EEG + fNIRS   & 0.2494 & 0.2280 & 0.2387 \\
		\midrule
		\multirow{3}{*}{ASAC-Net~\cite{FANG2026104329}}
		& EEG-only    &0.2613 & 0.2472 & 0.2543  \\
		& fNIRS-only   &0.2465 & 0.2254 & 0.2360  \\
		& EEG + fNIRS   & 0.2307 & 0.2020 & 0.2164  \\
		\bottomrule
	\end{tabular}
    }
\end{table}

\section{Challenge Guidance}

The MER2026 datasets are strictly for academic research purposes. Users are prohibited from modifying the source videos or uploading them to the internet. For \textbf{MER-Cross}, \textbf{MER-FG}, and \textbf{MER-Prefer}, we provide a list covering the test set for each track, requiring participants to submit predictions for all samples in the list. This process ensures that participants cannot directly access the test samples and engage in result exhaustion to achieve a higher ranking. For \textbf{MER-PS}, the local Institutional Review Board approved the ethical conduct of this study (IA11-2504-020201). All procedures involving human participants strictly adhered to established ethical standards, with particular emphasis on ensuring participant safety, privacy, and informed consent. Prior to participation, all subjects were fully informed about the study's purpose, procedures, and data usage policies. Thus, no ethical issues exist for this track.

\section{Conclusion}
\label{sec:conclusion}
MER2026 marks our fourth edition of the MER challenges. Centered on the theme ``From Discriminative Emotion Recognition to Generative Emotion Understanding'', this year's event aligns with current research trends. MER2026 comprises four tracks: MER-Cross focuses on interlocutor emotion recognition, MER-FG targets fine-grained emotion detection, MER-Prefer aims to mimic human emotion preference, and MER-PS introduces physiological signals for emotion recognition. This paper presents the task definition, dataset, metrics, and baselines for each track, providing participants with essential guidance. Let us work together to build a reliable and trustworthy future of affective computing systems.

\section*{Acknowledgement}
% We would like to express our sincere gratitude to the Dataset Chairs for their support during the dataset construction process: Xiaojiang Peng (Shenzhen Technology University), Kele Xu (National University of Defense Technology), Fei Ma (Guangdong Lab of Artificial Intelligence and Digital Economy (SZ)), Ziyu Jia (Institute of Automation, CAS), Laizhong Cui (Shenzhen University), and Zebang Cheng (Shenzhen University). We also thank the members of our annotation team for their efforts: Zelin Yan (Tongji University), Liyi Liu (Tongji University), Chenxi Zhou (National University of Defense Technology), Yuan Cao (National University of Defense Technology), Kaiyao Li (Shenzhen University), Dawei Huang (Shenzhen University), and Hanwen Du (Shenzhen University). Finally, we thank the ACM Multimedia 2026 Challenge Chairs for giving us the opportunity to organize MER2026 at this premier conference.

We would like to express our gratitude to our annotation team: Zelin Yan (Tongji University), Liyi Liu (Tongji University), Chenxi Zhou (National University of Defense Technology), Yuan Cao (National University of Defense Technology), Kaiyao Li (Shenzhen University), Dawei Huang (Shenzhen University), and Hanwen Du (Shenzhen University). We also thank the ACM Multimedia organizers and challenge chairs for their assistance. This work is supported by the Open Research Fund of the State Key Laboratory of Multimodal Artificial Intelligence Systems (MAIS2026003).

\newpage
\bibliographystyle{unsrt}
\bibliography{mybib}

\end{document}